\documentstyle[12pt]{article}

\let \ttorg \tt \def \tt{\ttorg \obeyspaces}

\begin{document}

\date{}

\title{\Large\bf Comparing Quantum Entanglement and Topological Entanglement} \author{Louis
H. Kauffman\\ Department of Mathematics, Statistics \\ and Computer Science (m/c
249)    \\ 851 South Morgan Street   \\ University of Illinois at Chicago\\
Chicago, Illinois 60607-7045\\ $<$kauffman@uic.edu$>$\\ and \\ Samuel J. Lomonaco
Jr. \\ Department of Computer Science and Electrical Engineering \\ University of
Maryland Baltimore County \\ 1000 Hilltop Circle, Baltimore, MD 21250\\
$<$lomonaco@umbc.edu$>$}

\maketitle

\thispagestyle{empty}

\subsection*{\centering Abstract}

{\em This paper discusses relationships between topological entanglement and
quantum entanglement. Specifically, we propose that it is more fundamental to
view topological entanglements such as braids as {\em entanglement operators} and
to associate to them unitary operators that are capable of creating quantum entanglement.}

\section{Introduction} This paper discusses relationships between topological
entanglemenet and quantum entanglement. The present paper is an 
expanded version of \cite{TEQE}.  Specifically, we propose that it is more
fundamental to view topological entanglements such as braids as {\em entanglement
operators} and to associate to them unitary operators that perform quantum
entanglement. Then one can compare the way the unitary operator corresponding to
an elementary braid has (or has not) the capacity to entangle quantum states.
Along with this, one can examine the capacity of the same operator to detect
linking. The detection of linking involves working with closed braids or with
link diagrams. In both cases, the algorithms for computing link invariants are
very interesting to examine in the light of quantum computing. These algorithms
can usually be decomposed into one part that is a straight composition of unitary
operators, and hence can be seen as a sequence of quantum computer instructions, and another part that can
be seen either as preparation/detection, or as a quantum network with cycles in the underlying graph.
\bigbreak

The paper is organized as follows. Section 2 discusses the basic analogy between
topological entanglement and quantum entanglement. Section 3 proposes the
viewpoint of braiding operators and gives a specific example of a unitary
braiding operator, showing that it does entangle quantum states. Section 3 ends
with a list of problems. Section 4 discusses the link invariants associated with
the braiding operator $R$ introduced in the previous section. Section 5 is a discussion of 
the structure of entanglement in relation to measurement.
Section 6 is an
introduction to the virtual braid group, an extension of the classical braid
group by the symmetric group. We contend that unitary representations of the
virtual braid group provide a good context and language for quantum computing. Section
7 is a discussion of ideas and concepts that have arisen in the course of this
research.
\bigbreak

\noindent {\bf Acknowledgement.} Most of this effort was sponsored by the Defense
Advanced Research Projects Agency (DARPA) and Air Force Research Laboratory, Air
Force Materiel Command, USAF, under agreement F30602-01-2-05022. Some of this
effort was also sponsored by the National Institute for Standards and Technology
(NIST). The U.S. Government is authorized to reproduce and distribute reprints
for Government purposes notwithstanding any copyright annotations thereon. The
views and conclusions contained herein are those of the authors and should not be
interpreted as necessarily representing the official policies or endorsements,
either expressed or implied, of the Defense Advanced Research Projects Agency,
the Air Force Research Laboratory, or the U.S. Government. (Copyright 2002.) It
gives the first author great pleasure to thank Fernando Souza for interesting
conversations in the course of preparing this paper. \bigbreak

\section{The Temptation of Tangled States} It is quite tempting to make an
analogy between topological entanglement in the form of linked loops in three
dimensional space and the entanglement of quantum states. A topological
entanglement is a non-local structural feature of a topological system. A quantum
entanglement is a non-local structural feature of a quantum system. Take the case
of the Hopf link of linking number one. See Figure 1.  In this Figure we show a
simple link of two components and state its inequivalence to the disjoint union
of two unlinked loops.  The analogy that one wishes to draw is with a state of
the form $$\psi = (|01> - |10>)/\sqrt{2}$$ which is {\em quantum entangled.}
That is, this state is not of the form $\psi_{1} \otimes \psi_{2} \in H \otimes H$ where $H$
is a complex vector space of dimension two. Cutting a component of the link removes
its topological entangement. Observing the state removes its quantum entanglement in this case.
\bigbreak

{\tt    \setlength{\unitlength}{0.92pt} \begin{picture}(282,127) \thicklines  
\put(240,45){\framebox(41,41){}} \put(191,45){\framebox(41,41){}}
\put(168,85){\line(-2,-3){21}} \put(166,69){\vector(-1,0){29}}
\put(150,69){\vector(1,0){24}} \put(78,84){\line(-1,0){34}}
\put(123,84){\line(-1,0){34}} \put(124,4){\line(0,1){80}}
\put(44,3){\line(1,0){80}} \put(44,84){\line(0,-1){81}}
\put(3,45){\line(0,1){78}} \put(36,44){\line(-1,0){32}}
\put(84,44){\line(-1,0){36}} \put(85,124){\line(0,-1){80}}
\put(3,124){\line(1,0){82}} \end{picture}}

{\bf Figure 1 - The Hopf Link} \bigbreak

An example of Aravind \cite{Ara} makes the possibility of such a connection even
more tantalizing. Aravind compares the Borommean Rings (See Figure 2) and the
$GHZ$ state $$|\psi> = (|\beta_{1}>|\beta_{2}>|\beta_{3}> -
|\alpha_{1}>|\alpha_{2}>|\alpha_{3}>)/\sqrt{2}.$$ \bigbreak

{\tt    \setlength{\unitlength}{0.92pt} \begin{picture}(145,141) \thicklines  
\put(61,138){\line(1,0){81}} \put(142,23){\line(0,1){114}}
\put(122,22){\line(1,0){19}} \put(61,22){\line(1,0){50}}
\put(61,75){\line(0,-1){53}} \put(61,138){\line(0,-1){51}}
\put(116,83){\line(-1,0){11}} \put(116,3){\line(0,1){80}}
\put(3,3){\line(1,0){113}} \put(3,82){\line(0,-1){80}}
\put(13,82){\line(-1,0){10}} \put(28,82){\line(1,0){62}}
\put(21,42){\line(0,1){79}} \put(53,42){\line(-1,0){32}}
\put(99,42){\line(-1,0){34}} \put(102,42){\line(0,0){0}}
\put(101,122){\line(0,-1){80}} \put(69,122){\line(1,0){32}}
\put(21,122){\line(1,0){32}} \end{picture}}

{\bf Figure 2 - Borommean Rings} \bigbreak

The Borommean Rings are a three component link with the property that the triplet
of components is indeed topologically linked, but the removal of any single
component leaves a pair of unlinked rings. Thus, the Borommean Rings are of
independent intellectual interest as an example of a tripartite relation that is
not expressed in terms of binary relations. The $GHZ$ state can be viewed as
an entangled superposition of three particles with (say) all their spins in the
$z-$ direction. If we measure one particle of the three particle quantum system, then the state becomes disentangled
(That is, it becomes a tensor product). Thus the $GHZ$ state appears to be a quantum analog to
the Borommean Rings! 
\bigbreak

However, Aravind points out that this analogy is basis dependent,
for if one changes basis, rewriting to 
$$|\psi> =
(|\beta_{1x}>/\sqrt{2})(|\beta_{2}>|\beta_{3}> -
|\alpha_{2}>|\alpha_{3}>)/\sqrt{2}$$ $$ +
(|\alpha_{1x}>/\sqrt{2})(|\beta_{2}>|\beta_{3}> +
|\alpha_{2}>|\alpha_{3}>)/\sqrt{2},$$

\noindent where $|\beta_{1x}>$ and $|\alpha_{1x}>$ denote the spin-up and
spin-down states of particle $1$ in the $x$ direction, then one sees that a
measurement of the spin of particle $1$ in the $x$ direction will yield an
entangled state of the other two particles. Thus, in this basis, the state $|\psi>$
behaves like a triplet of loops such that each pair of loops is linked! Seeing
the state as analogous to a specific link depends upon the choice of basis. From
a physical standpoint, seeing the state as analogous to a link depends upon the
choice of an observable. \bigbreak

These examples show that the analogy between topological linking and quantum
entanglement is surely complex. One might expect a collection of links to
exemplify the entanglement properties of a single quantum state. It is attractive
to consider the question: {\em What patterns of linking are inherent in a given
quantum state?} This is essentially a problem in linear algebra and  should be
investigated further. We will not pursue it in this paper. 
\bigbreak

On top of this, there
is quite a bit of ingenuity required to produce links with given properties. For
example, in Figure 3 we illustrate a Brunnian Link of four components. This link
has the same property as the Borommean Rings but for four components rather than
three. Remove any component and the link falls apart. The obvious generalization
of th $GHZ$ state with this property just involves adding one more tensor product
in the two-term formula. This raises a question about the relationship of
toplogical complexity and algebraic complexity of the corresponding quantum
state. The other difficulties with this analogy are that topological properties
of linked loops are not related to quantum mechanics in any clear way.
Nevertheless, it is clear that this is an analogy worth pursuing. \bigbreak

{\tt    \setlength{\unitlength}{0.92pt} \begin{picture}(367,329) \thicklines  
\put(18,59){\line(0,1){44}} \put(303,23){\line(-1,0){76}}
\put(204,3){\line(1,0){122}} \put(243,126){\line(0,-1){25}}
\put(126,189){\line(0,-1){24}} \put(170,83){\line(-1,0){25}}
\put(170,97){\line(0,-1){13}} \put(3,125){\line(0,-1){86}}
\put(198,40){\line(-1,0){196}} \put(222,40){\line(-1,0){13}}
\put(243,40){\line(-1,0){14}} \put(243,58){\line(0,-1){17}}
\put(242,59){\line(-1,0){223}} \put(53,103){\line(-1,0){35}}
\put(140,102){\line(-1,0){77}} \put(243,101){\line(-1,0){94}}
\put(53,126){\line(-1,0){49}} \put(139,126){\line(-1,0){76}}
\put(199,126){\line(-1,0){50}} \put(218,126){\line(-1,0){10}}
\put(242,126){\line(-1,0){14}} \put(204,54){\line(0,-1){49}}
\put(204,98){\line(0,-1){37}} \put(203,161){\line(0,-1){52}}
\put(204,184){\line(0,-1){14}} \put(203,205){\line(0,-1){13}}
\put(226,55){\line(0,-1){32}} \put(225,101){\line(0,-1){39}}
\put(223,206){\line(0,-1){98}} \put(326,160){\line(0,-1){155}}
\put(325,186){\line(0,-1){17}} \put(324,205){\line(0,-1){12}}
\put(141,165){\line(-1,0){15}} \put(163,165){\line(-1,0){13}}
\put(219,166){\line(-1,0){46}} \put(298,165){\line(-1,0){69}}
\put(363,165){\line(-1,0){55}} \put(220,189){\line(-1,0){93}}
\put(298,189){\line(-1,0){70}} \put(339,190){\line(-1,0){31}}
\put(171,274){\line(1,0){193}} \put(149,274){\line(1,0){12}}
\put(123,274){\line(1,0){17}} \put(169,110){\line(0,-1){2}}
\put(169,123){\line(0,-1){13}} \put(168,186){\line(0,-1){57}}
\put(167,246){\line(0,-1){53}} \put(166,326){\line(0,-1){72}}
\put(144,183){\line(0,-1){99}} \put(144,246){\line(0,-1){53}}
\put(144,302){\line(0,-1){47}} \put(32,85){\line(1,0){26}}
\put(57,303){\line(0,-1){218}} \put(32,100){\line(0,-1){14}}
\put(32,121){\line(0,-1){13}} \put(31,326){\line(0,-1){196}}
\put(303,206){\line(1,0){21}} \put(303,24){\line(0,1){181}}
\put(204,206){\line(1,0){18}} \put(339,251){\line(0,-1){61}}
\put(125,251){\line(1,0){214}} \put(123,273){\line(0,-1){22}}
\put(57,303){\line(1,0){86}} \put(32,326){\line(1,0){133}}
\put(364,274){\line(0,-1){109}} \end{picture}}

{\bf Figure 3 - A Brunnian Link} \bigbreak

\section{Entanglement Operators} Braids and the Artin braid group form a first
instance in topology where a space (or topological configuration) is also seen as
an {\em operator} on spaces and configurations. It is a shift that transmutes the
elements of a topological category to morphisms in an associated category. While
we shall concentrate on braids as an exemplar of this shift, it is worth noting
that such a shift is the basis of quantum topology and topological quantum field
theory, where spaces are viewed (through appropriate functors) as morphisms in a
category analogous to a category of Feynman diagrams. This pivot from spaces to
morphisms and back is the fundamental concept behind topological quantum field
theory. \bigbreak

Braids are patterns of entangled strings. A braid has the form of a collection of
strings extending from one set of points to another, with a constant number of
points in each cross section. Braids start in one row of points and end in
another. As a result, one can multiply two braids to form a third braid by
attaching the end points of the first braid to the initial points of the second
braid. Up to topological equivalence, this multiplication gives rise to a group,
the Artin Braid Group $B_{n}$ on $n$ strands. \smallbreak

Each braid is, in itself, a pattern of entanglement. Each braid is an operator
that operates on other patterns of entanglement (braids) to produce new
entanglements (braids again). \smallbreak

We wish to explore the analogy between topological entanglement and quantum
entanglement. From the point of view of braids this means {\em the association of
a unitary operator with a braid that repspects the topological structure of the
braid and allows exploration of the entanglement properties of the operator.} In
other words, we propose to study the analogy between topological entanglement and
quantum entanglement by looking at {\em unitary representations of the Artin
Braid Group}. It is not the purpose of this paper to give an exhaustive account
of such representations. Rather, we shall concentrate on one particularly simple
representation and analyze the relationships between topological and quantum
entanglement that are implicit in this representation. The main point for the
exploration of the analogy is that, from the point of view of a braid group
representation, each braid is seen as an operator rather than a state. See Figure
4. \bigbreak

{\tt    \setlength{\unitlength}{0.92pt} \begin{picture}(399,160) \thicklines  
\put(257,145){\line(1,0){80}} \put(256,105){\line(1,0){80}}
\put(256,65){\line(1,0){80}} \put(318,65){\line(1,0){80}}
\put(319,145){\line(1,0){18}} \put(337,145){\line(1,-1){39}}
\put(376,106){\line(1,0){22}} \put(318,105){\line(1,0){19}}
\put(337,105){\line(1,1){17}} \put(361,129){\line(1,1){16}}
\put(377,145){\line(1,0){21}} \put(130,49){\framebox(63,110){}}
\put(179,144){\line(1,0){21}} \put(163,128){\line(1,1){16}}
\put(139,104){\line(1,1){17}} \put(120,104){\line(1,0){19}}
\put(178,105){\line(1,0){22}} \put(139,144){\line(1,-1){39}}
\put(121,144){\line(1,0){18}} \put(107,1){\makebox(122,39){Braiding Operator}}
\put(214,106){\vector(1,0){30}} \put(120,64){\line(1,0){80}}
\put(1,63){\line(1,0){80}} \put(1,104){\line(1,0){80}}
\put(2,144){\line(1,0){80}} \end{picture}}

{\bf Figure 4  - A Braiding Operator} \bigbreak

{\tt    \setlength{\unitlength}{0.92pt}
\begin{picture}(282,364)
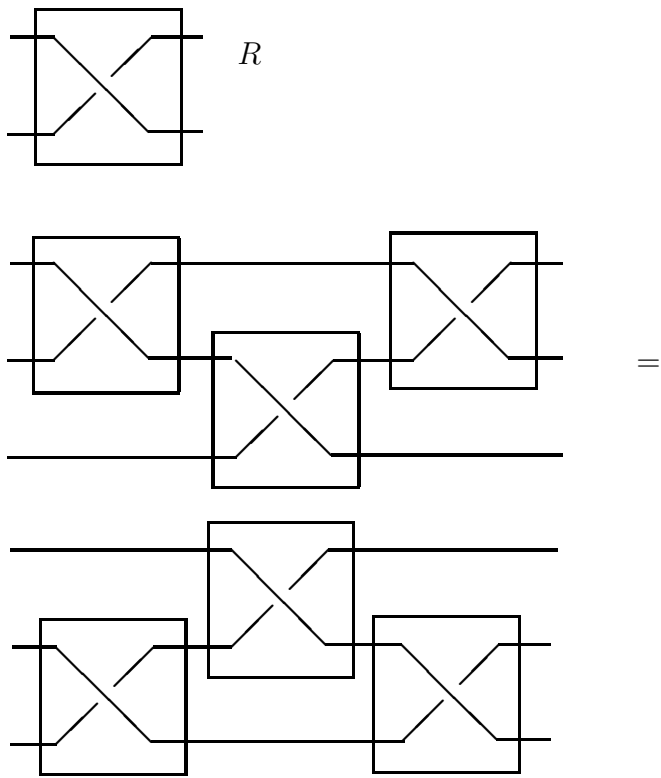
\thicklines   \put(39,1){\makebox(242,41){The Yang-Baxter Equation}}
              \put(88,333){\makebox(25,25){$R$}}
              \put(249,202){\makebox(31,30){$=$}}
              \put(152,50){\framebox(59,63){}}
              \put(84,89){\framebox(59,63){}}
              \put(15,49){\framebox(59,63){}}
              \put(159,208){\framebox(59,63){}}
              \put(86,167){\framebox(59,63){}}
              \put(12,206){\framebox(59,63){}}
              \put(13,300){\framebox(59,63){}}
              \put(72,62){\line(1,0){80}}
              \put(75,141){\line(1,0){18}}
              \put(93,141){\line(1,-1){39}}
              \put(132,102){\line(1,0){22}}
              \put(74,101){\line(1,0){19}}
              \put(93,101){\line(1,1){17}}
              \put(117,125){\line(1,1){16}}
              \put(133,141){\line(1,0){21}}
              \put(61,101){\line(1,0){21}}
              \put(45,85){\line(1,1){16}}
              \put(21,61){\line(1,1){17}}
              \put(2,61){\line(1,0){19}}
              \put(60,62){\line(1,0){22}}
              \put(21,101){\line(1,-1){39}}
              \put(3,101){\line(1,0){18}}
              \put(2,141){\line(1,0){80}}
              \put(203,102){\line(1,0){21}}
              \put(187,86){\line(1,1){16}}
              \put(163,62){\line(1,1){17}}
              \put(145,62){\line(1,0){19}}
              \put(202,63){\line(1,0){22}}
              \put(163,102){\line(1,-1){39}}
              \put(145,102){\line(1,0){18}}
              \put(147,141){\line(1,0){80}}
              \put(149,180){\line(1,0){80}}
              \put(150,259){\line(1,0){18}}
              \put(168,259){\line(1,-1){39}}
              \put(207,220){\line(1,0){22}}
              \put(149,219){\line(1,0){19}}
              \put(168,219){\line(1,1){17}}
              \put(192,243){\line(1,1){16}}
              \put(208,259){\line(1,0){21}}
              \put(2,352){\line(1,0){18}}
              \put(20,352){\line(1,-1){39}}
              \put(59,313){\line(1,0){22}}
              \put(1,312){\line(1,0){19}}
              \put(20,312){\line(1,1){17}}
              \put(44,336){\line(1,1){16}}
              \put(60,352){\line(1,0){21}}
              \put(76,259){\line(1,0){80}}
              \put(75,220){\line(1,0){18}}
              \put(95,219){\line(1,-1){39}}
              \put(134,180){\line(1,0){22}}
              \put(76,179){\line(1,0){19}}
              \put(95,179){\line(1,1){17}}
              \put(119,203){\line(1,1){16}}
              \put(135,219){\line(1,0){21}}
              \put(60,259){\line(1,0){21}}
              \put(44,243){\line(1,1){16}}
              \put(20,219){\line(1,1){17}}
              \put(1,219){\line(1,0){19}}
              \put(59,220){\line(1,0){22}}
              \put(20,259){\line(1,-1){39}}
              \put(2,259){\line(1,0){18}}
              \put(1,179){\line(1,0){80}}
\end{picture}}

{\bf Figure 5  - The Yang-Baxter Equation} \bigbreak

{\tt    \setlength{\unitlength}{0.92pt} \begin{picture}(367,65) \thicklines  
\put(256,1){\framebox(59,63){}} \put(168,23){\makebox(25,24){=}}
\put(13,1){\framebox(59,63){}} \put(206,14){\line(1,0){160}}
\put(206,54){\line(1,0){159}} \put(118,29){\line(1,-1){13}}
\put(95,54){\line(1,-1){17}} \put(95,15){\line(1,1){38}}
\put(134,53){\line(1,0){21}} \put(76,15){\line(1,0){19}}
\put(132,15){\line(1,0){22}} \put(75,54){\line(1,0){18}}
\put(85,1){\framebox(59,63){}} \put(2,54){\line(1,0){18}}
\put(20,54){\line(1,-1){39}} \put(59,15){\line(1,0){22}}
\put(1,14){\line(1,0){19}} \put(20,14){\line(1,1){17}}
\put(44,38){\line(1,1){16}} \put(60,54){\line(1,0){21}} \end{picture}}

{\bf Figure 6 - Inverses} \bigbreak

We will consider representations of the braid group constructed in the following
manner. To an elementary two strand braid there is associated an operator 
$$R: V \otimes V \longrightarrow V \otimes V.$$ Here $V$ is a complex vector space, and
for our purposes, $V$ will be two dimensional so that $V$ can hold a single qubit
of information. One should think of the two input and two output lines from the
braid as representing this map of tensor products. Thus the left endpoints of $R$
as shown in Figures 4, 5 and 6 represent the tensor product $V \otimes V$ that
forms the domain of $R$ and the right endpoints of the diagram for $R$ represent
$V \otimes V$ as the range of the maping. In the diagrams with three lines shown
in Figure 5, we have mappings from $V \otimes V \otimes V$ to itself. The identity
shown in Figure 5 is called the Yang-Baxter Equation, and it reads algebraically
as follows, where $I$ denotes the identity transformation on $V.$ $$(R \otimes
I)(I \otimes R)(R \otimes I)  =  (I \otimes R)(R \otimes I)(I \otimes R).$$
\noindent This equation expresses the fundamental topological relation in the
Artin Braid group, and is the main requirement for producing a representation of
the braid group by this method. We also need an inverse to $R$ and this will be
associated with the reversed elementary braid on two strands as shown in Figure
6. One then defines a representation $\tau$ of the Artin Braid Group to
automorphisms of $V^{\otimes n}$ by the equation $$\tau(\sigma_{k}) = I \otimes
\cdots \otimes I \otimes R \otimes I \cdots \otimes I,$$ 
\noindent where the $R$
occupies the $k$ and $k+1$ places in this tensor product. If $R$ satisfies the
Yang-Baxter equation and is invertible, then this formula describes a
representation of the braid group. If $R$ is unitary, then this construction
provides a unitary representation of the braid group. \bigbreak

Here is the specific $R$ matrix that we shall examine. The point of this case
study is that $R$, being unitary, can be considered as a quantum gate {\em and}
since $R$ is the key ingredient in a unitary representation of the braid group,
it can be considered as a operator that performs topological entanglement. We
shall see that it can also perform quantum entanglement in its action on quantum
states.

$$R = \left[ \begin{array}{cccc} a & 0 & 0 & 0 \\ 0 & 0 & d & 0 \\ 0 & c & 0 & 0
\\ 0 & 0 & 0 & b \end{array} \right].$$

\noindent Here $a,b,c,d$ can be any scalars on
the unit circle in the complex plane. {\em Then $R$ is a unitary matrix and it is
a solution to the Yang-Baxter Equation.} It is an interesting and illuminating
exercise to verify that $R$ is a solution to the Yang-Baxter Equation. We will
omit this verification here, but urge the reader to perform it. In fact, the
following more general construction gives a large class of unitary $R$ matrices:
Let $M = (M_{ij})$ denote an $n \times n$ matrix with entries in the unit circle
in the complex plane. Let $R$ be defined by the equation $$R^{ij}_{kl} =
\delta^{i}_{l} \delta^{j}_{k} M_{ij}.$$ \noindent It is easy to see that $R$ is a
unitary solution to the Yang-Baxter equation. Our explicit example is the special
case of $R$ where the matrix $M$ is $2 \times 2.$  It turns out, just as we shall
show here for the special case, $R$ detects no more than linking numbers for
braids, knots and links. This is interesting, but it would be even more
interesting to see other unitary $R$ matrices that have subtler topological
properties. The reader may enjoy comparing this situation with the unitary
representation of the Artin Braid Group discussed in \cite{QCJP}. 
\bigbreak

One can use
that representation to calculate the Jones polynomial for three-strand braids.
There is still a problem about designing a quantum computer to find the Jones
polynomial, but this braid group representation does encode subtle topology. At
the same time the representation in \cite{QCJP} cannot entangle quantum states. 
Thus the question of the precise relationship between
topological entanglement and quantum entanglement certainly awaits the arrival of
more examples of unitary representations of the braid group. We are indebted to
David Meyer for asking sharp questions in this domain \cite{DM}. \vspace{3mm}

\noindent Now let $P$ be the swap permutation matrix

$$P = \left[ \begin{array}{cccc} 1 & 0 & 0 & 0 \\ 0 & 0 & 1 & 0 \\ 0 & 1 & 0 & 0
\\ 0 & 0 & 0 & 1 \end{array} \right].$$

\noindent and let $\tau = RP$ so that

$$\tau = \left[ \begin{array}{cccc} a & 0 & 0 & 0 \\ 0 & c & 0 & 0 \\ 0 & 0 & d &
0 \\ 0 & 0 & 0 & b \end{array} \right].$$

\noindent Then from the point of view of quantum gates, we have the phase gate
$\tau$ and the swap gate $P$ with $\tau = RP$.  From the point of view of
braiding and algebra, we have that $R$ is a solution to the braided version of
the Yang-Baxter equation, $\tau$ is a solution to the algebraists version of the
Yang-Baxter equation, and $P$ is to be regarded as an algebraic permutation {\em
or} as a representation of a virtual or flat crossing. We discuss the virtual
braid group \cite{V1, V2, V3, Kam} in section 5, but for here suffice it to say
that it is an extension of the classical braid group by the symmetric group and
so contains braiding generators and also generators of order two. Now the point
is that by looking at unitary representations of the virtual braid group, we can
(as with the matrices above) pick up both phase and swap gates, and hence the
basic ingredients for quantum computation.  This means that the virtual braid
group provides a useful topological language for quantum computing. This deserves
further exploration. \vspace{3mm}

\noindent The matrix $R$ can also be used to make an invariant of knots and links
that is sensitive to linking numbers. We will discuss this point in section 4.
\vspace{3mm}

But now, consider the action of the unitary transformation $R$ on quantum states.
We have \begin{enumerate} \item $R|00> = a|00>$ \item $R|01> = c|10>$ \item
$R|10> = d|01>$ \item $R|11> = b|11>$ \end{enumerate} \bigbreak

\noindent Here is an elementary proof that the operator $R$ can entangle quantum states. Note
how this comes about through its being a composition of a phase and a swap gate.
This decomposition is available in the virtual braid group. \smallbreak

\noindent {\bf Lemma.} If $R$ is chosen so that $ab \ne cd$, then the state
$R(\psi \otimes \psi)$, with $\psi = |0> + |1>$, is entangled. \bigbreak

\noindent {\bf Proof.} By definition, $$\phi = R(\psi \otimes \psi) = R((|0> +
|1>) \otimes (|0> + |1>))$$ $$= a|00> + c|10> + d|01> + b|11>.$$ \noindent If
this state $\phi$ is unentangled, then there are constants $X$, $Y$, $X'$, $Y'$ such
that $$\phi = (X|0> + Y|1>) \otimes (X'|0> + Y'|1>).$$ \noindent This implies
that \begin{enumerate} \item $a = XX'$ \item $c = X'Y$ \item $d = XY'$ \item $b =
YY'$ \end{enumerate} \smallbreak

\noindent It follows from these equations that $ab = cd.$ Thus, when $ab \ne cd$
we can conclude that the state $\phi$ is entangled as a quantum state. //
\bigbreak

{\tt    \setlength{\unitlength}{0.92pt} \begin{picture}(373,77) \thicklines  
\put(251,18){\makebox(121,41){$\phi$  Entangled State}}
\put(2,1){\makebox(100,37){$|0> + |1>$}} \put(1,39){\makebox(100,37){$|0> +
|1>$}} \put(210,59){\vector(1,0){38}} \put(210,20){\vector(1,0){38}}
\put(102,19){\vector(1,0){38}} \put(102,59){\vector(1,0){38}}
\put(213,20){\line(1,0){19}} \put(212,59){\line(1,0){16}}
\put(139,59){\line(1,0){18}} \put(157,59){\line(1,-1){39}}
\put(196,20){\line(1,0){22}} \put(136,19){\line(1,0){19}}
\put(157,19){\line(1,1){17}} \put(181,43){\line(1,1){16}}
\put(197,59){\line(1,0){21}} \end{picture}}

{\bf Figure 7 - Braiding Operator Entangling a State} \bigbreak

\noindent {\bf Remark.} Note that if $\alpha = a|0> + b|1>$ and $\beta = c|0> + d|1>$
then $\alpha \otimes \beta = ac|00> + ad|01> + bc|10> + bd|11>.$ Thus a state
$\gamma = X|00> + Y|01> + Z|10> + W|11>$ is entangled if $XW \ne YZ.$
\bigbreak

\subsection{\bf Questions} This phenomenon leads to more questions than we have
answers. \begin{enumerate} \item How does one classify quantum entanglements in terms of
braids (and corresponding braiding operators) that can produce them. \item Can
all quantum entangled states be lifted to braidings? \item How do protocols for
quantum computing look from this braided point of view? \item What is the
relationship between the analogy between quantum states and entangled loops when
viewed through the lens of the braiding operators? \item Does the
association of unitary braiding operators shed light on quantum computing
algorithms for knot invariants and statistical mechanics models? Here one can
think of the computation of a knot invariant as separated into a braiding
computation that is indeed a quantum computation, plus an evaluation related to
the preparation and detection of a state(See \cite{KP, QCJP}). \item How does one classify all
unitary solutions to the Yang-Baxter equation. \end{enumerate} \bigbreak

\section {Link Invariants from R}

The unitary $R$ matrix that we have considered in this paper gives rise to a
non-trivial invariant of links. In this section we shall discuss the invariant
associated with the specialization of $R$ with $c=d$ so that

$$R = \left[ \begin{array}{cccc} a & 0 & 0 & 0 \\ 0 & 0 & c & 0 \\ 0 & c & 0 & 0
\\ 0 & 0 & 0 & b \end{array} \right].$$

\noindent Later we will specialize further so that $a=b.$
We omit the details here, and just give the formula for this invariant in
the form of a state summation. The invariant has the form 
$$Z_{K} = a^{-w(K)}(\sqrt{a/b})^{rot(K)} <K>,$$
\noindent where $w(K)$ is the sum of the crossing
signs of the oriented link $K$ and $rot(K)$ is the rotation number (or Whitney
degree) of the planar diagram for $K$. See Figure 8. The bracket $<K>$ is the unnormalized
state sum for the invariant. This state sum is defined through the equations
shown in Figure 8. \bigbreak

{\tt    \setlength{\unitlength}{0.92pt} \begin{picture}(303,273) \thicklines  
\put(171,43){\makebox(25,23){1/Q}} \put(250,39){\makebox(28,27){Q}}
\put(17,1){\makebox(89,28){$Q = \sqrt{b/a}$}} \put(92,43){\makebox(25,23){1/Q}}
\put(11,38){\makebox(28,27){Q}} \put(285,80){\makebox(17,17){1}}
\put(205,79){\makebox(18,18){0}} \put(126,81){\makebox(17,17){1}}
\put(45,80){\makebox(18,18){0}} \put(163,69){\vector(1,0){40}}
\put(283,69){\vector(0,1){39}} \put(244,69){\vector(1,0){39}}
\put(243,108){\vector(0,-1){39}} \put(282,109){\vector(-1,0){39}}
\put(203,69){\vector(0,1){39}} \put(163,109){\vector(0,-1){40}}
\put(202,109){\vector(-1,0){39}} \put(83,69){\vector(0,1){39}}
\put(123,69){\vector(-1,0){40}} \put(123,109){\vector(0,-1){40}}
\put(83,109){\vector(1,0){40}} \put(3,69){\vector(0,1){40}}
\put(43,69){\vector(-1,0){40}} \put(43,108){\vector(0,-1){39}}
\put(3,109){\vector(1,0){40}} \put(204,157){\makebox(41,24){$+ 1/c$}}
\put(123,160){\makebox(41,22){$+ 1/b$}} \put(43,160){\makebox(41,21){$= 1/a$}}
\put(27,167){\vector(1,-1){16}} \put(3,190){\vector(1,-1){16}}
\put(3,151){\vector(1,1){40}} \put(244,190){\vector(1,-1){39}}
\put(244,150){\vector(1,1){40}} \put(93,177){\line(0,-1){14}}
\put(99,177){\line(0,-1){14}} \put(268,190){\line(0,-1){14}}
\put(262,190){\line(0,-1){14}} \put(84,190){\vector(1,0){39}}
\put(84,150){\vector(1,0){40}} \put(165,149){\vector(1,0){40}}
\put(165,189){\vector(1,0){39}} \put(180,176){\line(0,-1){14}}
\put(174,176){\line(0,-1){14}} \put(258,183){\line(1,0){14}}
\put(102,159){\makebox(18,18){$0$}} \put(184,160){\makebox(17,17){$1$}}
\put(208,237){\makebox(28,24){$+ c$}} \put(125,237){\makebox(32,25){$+ b$}}
\put(45,237){\makebox(32,25){$=  a$}} \put(183,240){\makebox(17,17){$1$}}
\put(101,239){\makebox(18,18){$0$}} \put(257,263){\line(1,0){14}}
\put(173,256){\line(0,-1){14}} \put(179,256){\line(0,-1){14}}
\put(164,269){\vector(1,0){39}} \put(164,229){\vector(1,0){40}}
\put(83,230){\vector(1,0){40}} \put(83,270){\vector(1,0){39}}
\put(261,270){\line(0,-1){14}} \put(267,270){\line(0,-1){14}}
\put(98,257){\line(0,-1){14}} \put(92,257){\line(0,-1){14}}
\put(243,230){\vector(1,1){40}} \put(243,270){\vector(1,-1){39}}
\put(27,254){\vector(1,1){15}} \put(3,230){\vector(1,1){16}}
\put(3,270){\vector(1,-1){40}} \end{picture}}

{\bf Figure 8  - Formulas for the State Summation} \bigbreak

In this Figure, the first crossing is positive, the second negative. The first two
diagrammatic equations correspond to terms in the matrices $R$ and $R^{-1}$
respectively. Note that the glyphs in these equations are labeled with $0$ or
$1.$  The first two terms correspond to the action of $R$ on $|00>$ and on $|11>$
respectively. The third term refers to the fact that $R$ acts on $|01>$ and
$|10>$ in the same way (by multiplying by $c$). However, these equations are
interpreted for the state summation as instructions for forming local states on
the link diagram. A global state on the link diagram is a choice of replacement
for each crossing in the diagram so that it is either replaced by parallel arcs
(as in the first two terms of each equation) or by crossed arcs (as in the third
term of each equation). The local assignments of $0$ and $1$ on the arcs must fit
together compatibly in a global state. Thus in a global state one can think of
the $0$ and $1$ as qubits `circulating" around simple closed curves in the plane.
Each such state of circulation is measured in terms of the qubit type and the
sense of rotation. These are the evaluations of cycles indicated below the two
main equations for the state sum. Each cycle is assigned either $Q$ or $1/Q$
where $Q = \sqrt{b/a}.$ The state sum is the summation of evaluations of all of
the possible states of qubit circulation where each state is evaluated by the
product of weights $a$,$b$,$c$ (and their inverses) coming from the expansion
equations, multiplied by the porduct of the evaluations $Q$ or $1/Q$ of the
simple closed curves in the state. This completes a summary of the algorithm.
\bigbreak

There are many ways to construe a state summation such as this. One can arrange
the knot or link with respect to a given direction in the plane, and see the
calculation as a vacuum-vacuum amplitude in a toy quantum field theory \cite{KP}.
One can look directly at it as a generalized statistical mechanics state
summation as we described it above. One can write the link as a closed braid and
regard a major part of the calculation as a composition of unitary braiding
operators. In this picture, a good piece of the algorithm can be construed as 
quantum. We believe that algorithms of this type, inherent in the study
of so-called quantum link invariants, should be investigated more deeply from the
point of view of quantum computing. In particular, the point of view of the
algorithm as a sum over states of circulating qubits can be formalized,
and will be the subject of another
paper. \bigbreak

An example of a computation of this invariant is in order. In Figure 9 we show
the admissible states for a Hopf link (a simple link of two circles) where both
circles have the same rotation sense in the plane. We then see that if $H$
denotes the Hopf link, then $<H> = a^{2}Q^{2} + b^{2}Q^{-2} + 2c^{2}$ whence
$$Z_{H} = Q^{-2}<H> = a^{2} + b^{2}Q^{-4} +2c^{2}Q^{-2}.$$

\noindent From this it is easy to see that the invariant $Z$ detects the
linkedness of the Hopf link. In fact $Z$ cannot detect linkedness of links with
linking number equal to zero. For example, $Z$ cannot detect the linkedness of the 
Whitehead link shown in Figure 10.\bigbreak

{\tt    \setlength{\unitlength}{0.92pt} \begin{picture}(286,287) \thinlines   
\put(160,195){\makebox(42,42){$H$}} \thicklines   \put(250,80){\line(1,-1){18}}
\put(175,40){\line(1,-1){18}} \put(195,38){\line(-1,-1){17}}
\put(191,41){\line(-1,-1){17}} \put(270,78){\line(-1,-1){17}}
\put(266,81){\line(-1,-1){17}} \put(241,44){\vector(-1,0){77}}
\put(203,85){\vector(1,0){80}} \put(162,123){\vector(1,0){81}}
\put(243,123){\vector(0,-1){81}} \put(163,43){\vector(0,1){80}}
\put(283,83){\vector(0,-1){80}} \put(283,3){\vector(-1,0){80}}
\put(203,3){\vector(0,1){80}} \put(44,42){\line(-1,-1){17}}
\put(40,45){\line(-1,-1){17}} \put(90,95){\line(-1,-1){17}}
\put(85,99){\line(-1,-1){17}} \put(86,48){\vector(-1,0){37}}
\put(48,49){\vector(0,1){36}} \put(43,4){\vector(-1,1){38}}
\put(82,123){\vector(1,-1){41}} \put(50,85){\vector(1,-1){37}}
\put(2,124){\vector(1,0){81}} \put(3,44){\vector(0,1){80}}
\put(123,84){\vector(0,-1){80}} \put(123,4){\vector(-1,0){80}}
\put(44,164){\vector(0,1){80}} \put(124,164){\vector(-1,0){80}}
\put(124,244){\vector(0,-1){80}} \put(90,244){\vector(1,0){34}}
\put(44,244){\vector(1,0){35}} \put(4,204){\vector(0,1){80}}
\put(38,204){\vector(-1,0){34}} \put(84,203){\vector(-1,0){33}}
\put(84,284){\vector(0,-1){81}} \put(3,284){\vector(1,0){81}} \end{picture}}

{\bf Figure 9   - States for the Hopf Link $H$} \bigbreak

{\tt    \setlength{\unitlength}{0.92pt}
\begin{picture}(233,326)
\thicklines   \put(140,81){\vector(1,0){91}}
              \put(160,162){\vector(-1,0){20}}
              \put(230,162){\vector(-1,0){20}}
              \put(65,123){\makebox(40,42){$W$}}
              \put(204,77){\vector(0,-1){74}}
              \put(204,323){\vector(0,-1){239}}
              \put(83,285){\vector(0,1){39}}
              \put(83,243){\vector(0,1){34}}
              \put(198,162){\vector(-1,0){30}}
              \put(164,87){\vector(0,1){155}}
              \put(163,42){\vector(0,1){31}}
              \put(123,235){\vector(0,-1){34}}
              \put(122,281){\vector(0,-1){34}}
              \put(230,82){\vector(0,1){80}}
              \put(139,162){\vector(0,-1){80}}
              \put(3,2){\vector(0,1){279}}
              \put(202,3){\vector(-1,0){199}}
              \put(83,323){\vector(1,0){121}}
              \put(164,243){\vector(-1,0){82}}
              \put(43,42){\vector(1,0){120}}
              \put(42,201){\vector(0,-1){158}}
              \put(123,201){\vector(-1,0){81}}
              \put(3,281){\vector(1,0){120}}
\end{picture}}

{\bf Figure 10  - The Whitehead Link} \bigbreak

\subsection{A Further Specialization of $Z_{K}$} If we let $a=b$ in the
definition of $Z_{K}$, then the state summation becomes particularly simple with
$Q=1.$ It is then easy to see that for a two component link $Z_{K}$ is given by
the formula $$Z_{K} = 2(1 + (c^{2}/a^{2})^{lk(K)})$$ \noindent where $lk(K)$
denotes the linking number of the two components of $K.$ Thus we see that {\em for
this specialization of the $R$ matrix the operator $R$ entangles quantum states
exactly when it can detect linking numbers in the topological context.}
\bigbreak

Here is another description of the state sum: Instead of smoothing or flattening the 
crossings of the diagram, label each component of the diagram with either $0$ or $1$.
Take vertex weights of $a$ or $c$ (in this special case, and the corresponding matrix
entries in the general case) for each local labelling of a positive crossing as shown in Figure 11.
For a negative crossing the corresponding labels are $1/a$ and $1/c$ (which are the complex 
conjugates of $a$ and $c$ repsectively, when $a$ and $c$ are unit complex numbers).  
Let each state (labelling of the diagram by zeroes and ones) contribute the product of its 
vertex weights. Let $\Sigma(K)$ denote the sum over all the states of the products of the vertex
weights. Then one can verify that $Z(K) = a^{-w(K)} \Sigma(K)$ where $w(K)$ is the sum of the crossing
signs of the diagram $K.$

{\tt    \setlength{\unitlength}{0.92pt}
\begin{picture}(202,203)
\thicklines   \put(173,21){\makebox(15,18){0}}
              \put(27,51){\makebox(15,18){0}}
              \put(26,173){\makebox(15,18){0}}
              \put(56,141){\makebox(15,18){0}}
              \put(142,52){\makebox(19,19){1}}
              \put(143,173){\makebox(19,19){1}}
              \put(176,139){\makebox(19,19){1}}
              \put(56,21){\makebox(19,19){1}}
              \put(59,179){\makebox(23,23){a}}
              \put(181,60){\makebox(20,21){c}}
              \put(61,61){\makebox(20,21){c}}
              \put(178,178){\makebox(23,23){a}}
              \put(120,41){\vector(1,0){79}}
              \put(160,1){\vector(0,1){34}}
              \put(160,49){\vector(0,1){32}}
              \put(2,42){\vector(1,0){79}}
              \put(42,2){\vector(0,1){34}}
              \put(41,50){\vector(0,1){32}}
              \put(121,160){\vector(1,0){79}}
              \put(161,120){\vector(0,1){34}}
              \put(160,168){\vector(0,1){32}}
              \put(40,169){\vector(0,1){32}}
              \put(41,121){\vector(0,1){34}}
              \put(1,161){\vector(1,0){79}}
\end{picture}}

{\bf Figure 11 - Positive Crossing Weights}
\bigbreak

{\tt    \setlength{\unitlength}{0.92pt}
\begin{picture}(290,287)
\thicklines   \put(28,51){\makebox(20,22){0}}
              \put(101,86){\makebox(20,22){0}}
              \put(216,182){\makebox(20,22){0}}
              \put(258,63){\makebox(17,19){1}}
              \put(191,52){\makebox(17,19){1}}
              \put(221,23){\makebox(17,19){1}}
              \put(68,92){\makebox(17,19){1}}
              \put(60,24){\makebox(17,19){1}}
              \put(231,92){\makebox(17,19){1}}
              \put(258,245){\makebox(17,19){1}}
              \put(95,220){\makebox(20,22){0}}
              \put(224,249){\makebox(20,22){0}}
              \put(63,251){\makebox(20,22){0}}
              \put(59,181){\makebox(20,22){0}}
              \put(186,212){\makebox(17,19){1}}
              \put(26,212){\makebox(20,22){0}}
              \put(208,84){\vector(1,0){80}}
              \put(287,83){\vector(0,-1){80}}
              \put(288,3){\vector(-1,0){80}}
              \put(247,45){\vector(0,1){32}}
              \put(247,90){\vector(0,1){34}}
              \put(246,123){\vector(-1,0){81}}
              \put(165,124){\vector(0,-1){79}}
              \put(167,45){\vector(1,0){80}}
              \put(208,3){\vector(0,1){36}}
              \put(208,51){\vector(0,1){32}}
              \put(47,85){\vector(1,0){80}}
              \put(126,84){\vector(0,-1){80}}
              \put(127,4){\vector(-1,0){80}}
              \put(86,46){\vector(0,1){32}}
              \put(86,91){\vector(0,1){34}}
              \put(85,124){\vector(-1,0){81}}
              \put(4,125){\vector(0,-1){79}}
              \put(6,46){\vector(1,0){80}}
              \put(47,4){\vector(0,1){36}}
              \put(47,52){\vector(0,1){32}}
              \put(46,244){\vector(1,0){80}}
              \put(125,243){\vector(0,-1){80}}
              \put(126,163){\vector(-1,0){80}}
              \put(85,205){\vector(0,1){32}}
              \put(85,250){\vector(0,1){34}}
              \put(84,283){\vector(-1,0){81}}
              \put(3,284){\vector(0,-1){79}}
              \put(5,205){\vector(1,0){80}}
              \put(46,163){\vector(0,1){36}}
              \put(46,211){\vector(0,1){32}}
              \put(205,212){\vector(0,1){32}}
              \put(205,164){\vector(0,1){36}}
              \put(164,206){\vector(1,0){80}}
              \put(162,285){\vector(0,-1){79}}
              \put(243,284){\vector(-1,0){81}}
              \put(244,251){\vector(0,1){34}}
              \put(244,206){\vector(0,1){32}}
              \put(285,164){\vector(-1,0){80}}
              \put(284,244){\vector(0,-1){80}}
              \put(205,245){\vector(1,0){80}}
\end{picture}}

{\bf Figure 12  - Zero-One States for the Hopf Link} \bigbreak

For example, view Figure 12. Here we show the zero-one states for the Hopf link $H$.
The $00$ and $11$ states each contributes $a^2,$ while the $01$ and $10$ states contribute
$c^2.$ Hence $\Sigma(H) = 2(a^{2} + c^{2})$ and $a^{-w(H)}\Sigma(H) = 
2(1 + (c^{2}/a^{2})^{1}) = 2(1 + (c^{2}/a^{2})^{lk(H)}),$ as expected.
\bigbreak

The 
calculation of the invariant in this form is actually an analysis of quantum networks with cycles in the
underlying graph. In this form of calculation we are concerned with those states of the network
that correspond to 
labelings by qubits that are compatible with the entire network structure. A precise definition
of this concept will be given in a sequel to this paper. Here one considers only
those quantum states that are compatible with the interconnectedness of the network as a whole.
\smallbreak

The example of the Hopf link
shows how subtle properties of topological entanglement are detected through
the use of the operator $R$ in circularly interconnected quantum networks. It
remains to do a deeper analysis that can really begin to disentangle the roles of
quantum entanglement and circularity in such calculations. \bigbreak

\section{A Remark about EPR} It is remarkable that the simple algebraic situation
of an element in a tensor product that is not itself a a tensor product of
elements of the factors corresponds to subtle nonlocality in physics. It helps to
place this algebraic structure in the context of a gedanken experiment to see
where the physics comes in. Consider $$S = |0>|1> + |1>|0>.$$

\noindent
In an EPR thought experiment, we think of two ``parts" of this state that are
separated in space.  We want a notation for these parts and suggest the following:

$$L = \{|0>\}|1> + \{|1>\}|0>,$$

$$R = |0>\{|1>\} + |1>\{|0>\}.$$

\noindent In the left
state  $L$, an observer can only observe the left hand factor. In the right state $R$,
an observer can only observe the right hand factor.
These ``states"  $L$ and $R$ together comprise the EPR state $S,$ but they are
accessible individually just as are the two photons in the usual thought
experiement.  One can transport $L$ and $R$ individually and we shall write

$$S = L*R$$

\noindent to denote that they are the ``parts"  (but not tensor factors) of $S.$
\bigbreak

The curious thing about this formalism is that it includes a little bit of
macroscopic physics implicitly, and so it makes it a bit more apparent what EPR
were concerned about.  After all, lots of things that we can do to $L$ or $R$ do not
affect $S.$ For example, transporting $L$ from one place to another, as in the
original experiment where the photons separate.  On the other hand, if Alice has
$L$ and Bob has $R$ and Alice performs a local unitary transformation on ``her" tensor
factor, this applies to both $L$ and $R$ since the transformation is actually being
applied to the state $S.$ This is also a ``spooky action at a distance" whose
consequence does not appear until a measurement is made.
\bigbreak

\section {Virtual Braids} This section expands the remarks about how the
inclusion of a swap operator in the braid group leads to a significant
generalization of that structure to the virtual braid group. 
\bigbreak

The {\em virtual
braid group} is an extension of the classical braid group by the symmetric group.
If $V_{n}$ denotes the $n$--strand virtual braid group, then $V_{n}$ is generated
by  braid generators $\sigma_{1}$, ..., $\sigma_{n-1}$ and virtual generators
$c_{1}$,..., $c_{n}$ where each virtual generator $c_{i}$ has the form of the
braid generator $\sigma_{i}$ with the crossing replaced by a virtual crossing.
Among themselves, the braid generators satisfy the usual braiding relations. Among
themselves, the virtual generators are a presentation for the symmetric group
$S_{n}.$  The relations that relate virtual generators and braiding geneerators
are as follows:

$$\sigma_{i}^{\pm}c_{i+1}c_{i} = c_{i+1}c_{i} \sigma_{i+1}^{\pm},$$
$$c_{i}c_{i+1} \sigma_{i}^{\pm} = \sigma_{i+1}^{\pm}c_{i}c_{i+1},$$ $$c_{i}
\sigma_{i+1}^{\pm}c_{i} = c_{i+1} \sigma_{i}^{\pm}c_{i+1}.$$ \vspace{3mm}

\noindent It is easy to see from this description of the virtual braid groups
that all the braiding generators can be expressed in terms of the first braiding
generator $\sigma_{1}$ (and its inverse) and the virtual generators. One can also
see that Alexander's Theorem generalizes to virtuals: Every virtual knot is
equivalent to a virtual braid \cite{V2}.  In \cite{Kam} a Markov Theorem is
proven for virtual braids. \vspace{3mm}

{\tt    \setlength{\unitlength}{0.92pt} \begin{picture}(306,403) \thicklines  
\put(3,81){\line(0,-1){40}} \put(4,121){\line(1,-1){38}}
\put(3,82){\line(1,1){16}} \put(27,106){\line(1,1){16}}
\put(82,122){\line(0,-1){41}} \put(42,83){\line(1,-1){42}}
\put(82,80){\line(-1,-1){14}} \put(43,42){\line(1,1){16}}
\put(3,42){\line(1,-1){40}} \put(42,41){\line(-1,-1){15}}
\put(3,2){\line(1,1){17}} \put(83,40){\line(0,-1){38}}
\put(123,122){\line(0,-1){40}} \put(163,122){\line(1,-1){41}}
\put(202,122){\line(-1,-1){14}} \put(163,82){\line(1,1){16}}
\put(124,82){\line(1,-1){39}} \put(162,81){\line(-1,-1){15}}
\put(123,42){\line(1,1){17}} \put(204,81){\line(0,-1){39}}
\put(163,42){\line(1,-1){40}} \put(203,43){\line(-1,-1){16}}
\put(163,2){\line(1,1){17}} \put(124,42){\line(0,-1){40}}
\put(110,61){\vector(-1,0){16}} \put(93,61){\vector(1,0){19}}
\put(228,232){\framebox(77,50){$cc=1$}} \put(3,140){\framebox(201,49){$\sigma
\sigma^{-1} = 1$}} \put(221,359){\framebox(41,42){$c$}}
\put(56,306){\framebox(107,27){$\sigma^{-1}$}}
\put(56,360){\framebox(83,28){$\sigma$}} \put(134,240){\vector(1,0){19}}
\put(151,240){\vector(-1,0){16}} \put(54,241){\vector(1,0){19}}
\put(71,241){\vector(-1,0){16}} \put(184,380){\circle{22}}
\put(163,361){\line(1,1){39}} \put(164,400){\line(1,-1){40}}
\put(43,305){\line(-1,1){15}} \put(3,345){\line(1,-1){17}}
\put(43,345){\line(-1,-1){40}} \put(44,401){\line(-1,-1){16}}
\put(3,361){\line(1,1){18}} \put(4,400){\line(1,-1){40}}
\put(185,217){\circle{24}} \put(184,259){\circle{24}}
\put(162,240){\line(1,-1){40}} \put(203,281){\line(-1,-1){41}}
\put(204,238){\line(-1,-1){40}} \put(163,279){\line(1,-1){41}}
\put(123,281){\line(0,-1){79}} \put(82,281){\line(0,-1){82}}
\put(43,200){\line(-1,1){15}} \put(4,240){\line(1,-1){17}}
\put(43,240){\line(-1,-1){39}} \put(5,242){\line(1,1){17}}
\put(43,280){\line(-1,-1){14}} \put(4,281){\line(1,-1){40}} \end{picture}}

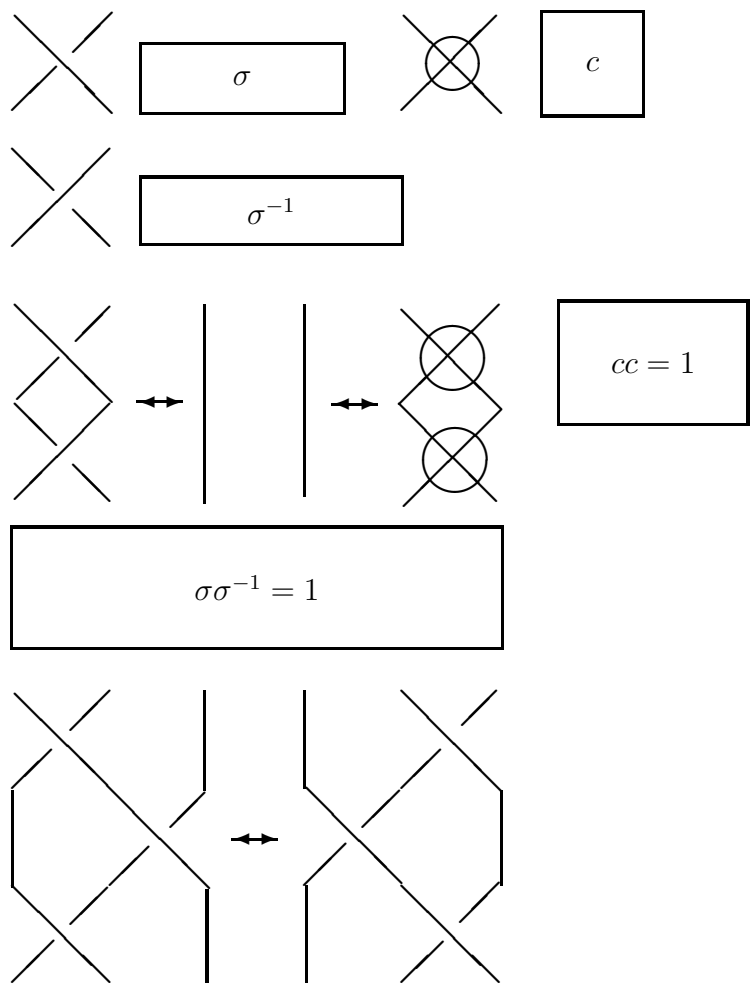
\noindent {\bf Figure 12 - Braid Generators and Virtual Braid Generators}
\vspace{3mm}

{\tt    \setlength{\unitlength}{0.92pt} \begin{picture}(208,445) \thicklines  
\put(183,182){\circle{22}} \put(145,221){\circle{22}} \put(63,221){\circle{22}}
\put(24,260){\circle{22}} \put(-54,461){\circle{0}}
\put(43,282){\line(-1,-1){39}} \put(203,203){\line(-1,-1){40}}
\put(124,202){\line(1,1){39}} \put(43,202){\line(1,1){39}}
\put(94,383){\vector(1,0){19}} \put(111,383){\vector(-1,0){16}}
\put(94,64){\vector(1,0){19}} \put(111,64){\vector(-1,0){16}}
\put(93,221){\vector(1,0){19}} \put(110,221){\vector(-1,0){16}}
\put(185,420){\circle{24}} \put(146,381){\circle{24}} \put(23,341){\circle{24}}
\put(63,383){\circle{24}} \put(124,362){\line(1,1){77}}
\put(2,323){\line(1,1){79}} \put(4,442){\line(1,-1){38}}
\put(3,403){\line(1,1){16}} \put(27,427){\line(1,1){16}}
\put(82,443){\line(0,-1){41}} \put(42,404){\line(1,-1){42}}
\put(3,402){\line(0,-1){39}} \put(3,363){\line(1,-1){40}}
\put(83,361){\line(0,-1){38}} \put(123,443){\line(0,-1){40}}
\put(163,443){\line(1,-1){41}} \put(124,403){\line(1,-1){39}}
\put(204,402){\line(0,-1){39}} \put(163,363){\line(1,-1){40}}
\put(203,364){\line(-1,-1){16}} \put(163,323){\line(1,1){17}}
\put(124,363){\line(0,-1){40}} \put(25,101){\circle{24}}
\put(185,22){\circle{24}} \put(185,101){\circle{24}} \put(26,21){\circle{24}}
\put(204,43){\line(-1,-1){39}} \put(164,83){\line(1,1){40}}
\put(4,2){\line(1,1){40}} \put(5,84){\line(1,1){39}} \put(5,122){\line(1,-1){38}}
\put(83,123){\line(0,-1){41}} \put(43,84){\line(1,-1){42}}
\put(83,81){\line(-1,-1){14}} \put(44,43){\line(1,1){16}}
\put(4,82){\line(0,-1){39}} \put(4,43){\line(1,-1){40}}
\put(84,41){\line(0,-1){38}} \put(124,123){\line(0,-1){40}}
\put(164,122){\line(1,-1){41}} \put(125,83){\line(1,-1){39}}
\put(163,82){\line(-1,-1){15}} \put(124,43){\line(1,1){17}}
\put(205,82){\line(0,-1){39}} \put(164,43){\line(1,-1){40}}
\put(125,43){\line(0,-1){40}} \put(124,202){\line(0,-1){40}}
\put(163,202){\line(1,-1){40}} \put(204,241){\line(0,-1){39}}
\put(124,242){\line(1,-1){39}} \put(163,242){\line(1,1){16}}
\put(202,282){\line(-1,-1){14}} \put(163,282){\line(1,-1){41}}
\put(123,282){\line(0,-1){40}} \put(83,200){\line(0,-1){38}}
\put(3,162){\line(1,1){17}} \put(42,201){\line(-1,-1){15}}
\put(3,202){\line(1,-1){40}} \put(3,241){\line(0,-1){39}}
\put(42,243){\line(1,-1){42}} \put(82,282){\line(0,-1){41}}
\put(4,281){\line(1,-1){38}} \end{picture}}

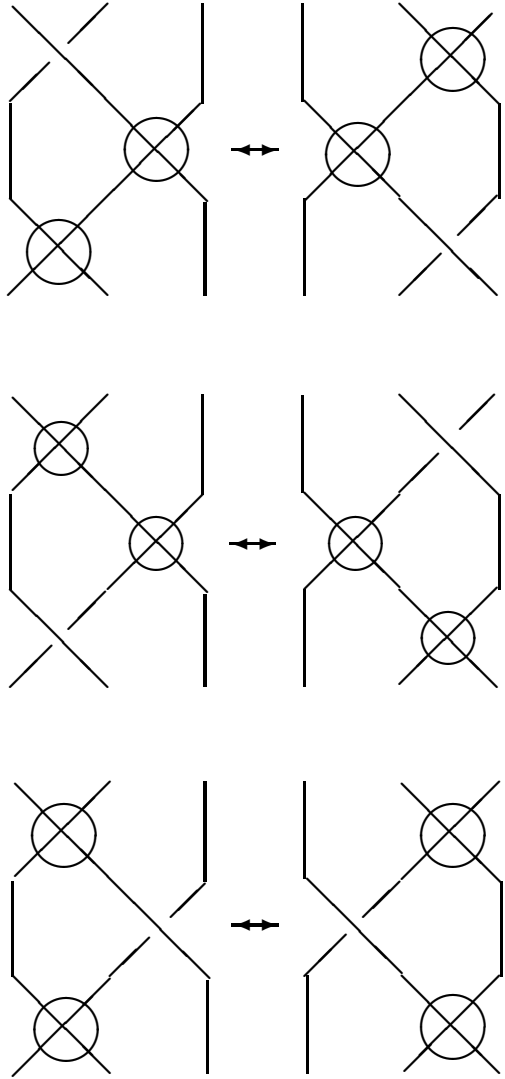
\noindent {\bf Figure 13 - Relations in the Virtual Braid Group} \vspace{3mm}

From the point of view of quantum computing, it is natural to add the virtual
braiding operators to the Artin Braid Group. {\em Each virtual braiding operator
can be interpreted as a swap gate.}  With the virtual operators in place, we can
compose them with the $R$ matrices to obtain phase gates and other apparatus as
described in Section 3. We then have the virtual braid group as a natural
topologically based group structure that can be used as an underlying language
for building patterns of quantum computation. \bigbreak

\section{Discussion} We are now in a position to state the main problem posed by
this paper.  We have been exploring the analogy between topological entanglement
and quantum entanglement. It has been suggested that there may be a direct
connection between these two phenomena. But on closer examination, it appears that
rather than a direct connection, there is a series of analogous features that are
best explored by going back and forth across the boundary between topology and
quantum computing. In particular, we have seen that the unitary operator $R$ can
indeed produce entangled quantum states from unentangled quantum states.  The
operator $R$ is the basic ingredient for forming a representation of the Artin
Braid Group. As such, it is intimately connected with topological entanglement. In
fact, the operator $R$ is also the basic ingredient in constructing the link
invariant $Z_{K}$ that we have studied in section 4. The construction of this
link invariant is motivated by quantum statistical mechanics and its structure
bears further investigation from the point of view of quantum computing. The
theme that emerges is powerfully related to the circularity of the links. It is
through mutual circularity that the topological linking occurs. And it is through
this circularity and the measurement of circulating states of qubits that one
computes the state summation model. A deep relation of quantum states and
topological states will be seen  through the study of the quantum states of
circularly interconnected networks structurally related to three-dimensional
space. These networks are both topological and quantum mechanical, and a common
structure will emerge. This is the project for further papers in our series.
\bigbreak

In the meantime, the language of the braid group and virtual braid group provides
an arena for representing quantum operators that can be interpreted
topologically. This framework provides a means for topology and quantum computing
to converse with one another. \bigbreak

The papers \cite{L,LO} and \cite{LP1,LP2} provide background to the considerations of the present paper.
In particular, they provide a general framework for studying quantum entanglement that 
may be useful in investigating the role of infinitesimal braiding operators and other aspects 
of the representation theory of the Artin braid group.
\bigbreak

The reader may wish to compare the points of view in this paper with the paper
\cite{F}. There the author considers the possibility of anyonic computing and
follows out the possible consequences in terms of representations of the Artin
Braid Group. We are in substantial agreement with his point of view {\em and} we
contend that braiding is fundamental to quantum computation whether or not it is
based in anyonic physics. \bigbreak

 \end{document}